\documentclass[10pt,twocolumn,twoside]{IEEEtran}

\usepackage{amssymb}
\usepackage{amsfonts}
\usepackage{amsmath}
\usepackage{cite}
\usepackage[dvips]{graphicx}
\usepackage{color}
\usepackage{comment}
\usepackage{makecell}
\usepackage{epstopdf}
\usepackage{gensymb}
\usepackage{subfigure}
\usepackage{algorithm}
\usepackage{algorithmic}

\begin{document}
\newtheorem{definition}{\it Definition}
\newtheorem{theorem}{\bf Theorem}
\newtheorem{lemma}{\it Lemma}
\newtheorem{corollary}{\it Corollary}
\newtheorem{remark}{\it Remark}
\newtheorem{example}{\it Example}
\newtheorem{case}{\bf Case Study}
\newtheorem{assumption}{\it Assumption}
\newtheorem{property}{\it Property}
\newtheorem{proposition}{\it Proposition}

\newcommand{\hP}[1]{{\boldsymbol h}_{{#1}{\bullet}}}
\newcommand{\hS}[1]{{\boldsymbol h}_{{\bullet}{#1}}}

\newcommand{\ba}{\boldsymbol{a}}
\newcommand{\baq}{\overline{q}}
\newcommand{\bA}{\boldsymbol{A}}
\newcommand{\bb}{\boldsymbol{b}}
\newcommand{\bB}{\boldsymbol{B}}
\newcommand{\bc}{\boldsymbol{c}}
\newcommand{\bcO}{\boldsymbol{\cal O}}
\newcommand{\be}{\boldsymbol{e}}
\newcommand{\bh}{\boldsymbol{h}}
\newcommand{\bH}{\boldsymbol{H}}
\newcommand{\bl}{\boldsymbol{l}}
\newcommand{\bm}{\boldsymbol{m}}
\newcommand{\bn}{\boldsymbol{n}}
\newcommand{\bo}{\boldsymbol{o}}
\newcommand{\bO}{\boldsymbol{O}}
\newcommand{\bp}{\boldsymbol{p}}
\newcommand{\bq}{\boldsymbol{q}}
\newcommand{\bR}{\boldsymbol{R}}
\newcommand{\bs}{\boldsymbol{s}}
\newcommand{\bS}{\boldsymbol{S}}
\newcommand{\bT}{\boldsymbol{T}}
\newcommand{\bw}{\boldsymbol{w}}

\newcommand{\balpha}{\boldsymbol{\alpha}}
\newcommand{\bbeta}{\boldsymbol{\beta}}
\newcommand{\bOmega}{\boldsymbol{\Omega}}
\newcommand{\bTheta}{\boldsymbol{\Theta}}
\newcommand{\bphi}{\boldsymbol{\phi}}
\newcommand{\btheta}{\boldsymbol{\theta}}
\newcommand{\bvarpi}{\boldsymbol{\varpi}}
\newcommand{\bpi}{\boldsymbol{\pi}}
\newcommand{\bpsi}{\boldsymbol{\psi}}
\newcommand{\bxi}{\boldsymbol{\xi}}
\newcommand{\bx}{\boldsymbol{x}}
\newcommand{\by}{\boldsymbol{y}}

\newcommand{\cA}{{\cal A}}
\newcommand{\bcA}{\boldsymbol{\cal A}}
\newcommand{\cB}{{\cal B}}
\newcommand{\cE}{{\cal E}}
\newcommand{\cG}{{\cal G}}
\newcommand{\cH}{{\cal H}}
\newcommand{\bcH}{\boldsymbol {\cal H}}
\newcommand{\cK}{{\cal K}}
\newcommand{\cO}{{\cal O}}
\newcommand{\cR}{{\cal R}}
\newcommand{\cS}{{\cal S}}
\newcommand{\dcS}{\ddot{{\cal S}}}
\newcommand{\ds}{\ddot{{s}}}
\newcommand{\cT}{{\cal T}}
\newcommand{\cU}{{\cal U}}
\newcommand{\wt}[1]{\widetilde{#1}}

\newcommand{\mA}{\mathbb{A}}
\newcommand{\mE}{\mathbb{E}}
\newcommand{\mG}{\mathbb{G}}
\newcommand{\mR}{\mathbb{R}}
\newcommand{\mS}{\mathbb{S}}
\newcommand{\mU}{\mathbb{U}}
\newcommand{\mV}{\mathbb{V}}
\newcommand{\mW}{\mathbb{W}}

\newcommand{\uq}{\underline{q}}
\newcommand{\ubq}{\underline{\boldsymbol q}}

\newcommand{\red}[1]{\textcolor[rgb]{1,0,0}{#1}}
\newcommand{\gre}[1]{\textcolor[rgb]{0,1,0}{#1}}
\newcommand{\blu}[1]{\textcolor[rgb]{0,0,1}{#1}}

\title{Optimizing Resource-Efficiency for Federated Edge Intelligence in IoT Networks} 

\author{\IEEEauthorblockA{Yong~Xiao\IEEEauthorrefmark{1}, Yingyu Li\IEEEauthorrefmark{1}, Guangming~Shi\IEEEauthorrefmark{2}, and H. Vincent Poor\IEEEauthorrefmark{3} \\
\IEEEauthorblockA{\IEEEauthorrefmark{1}School of Elect. Inform. \& Commun., Huazhong Univ. of Science \& Technology, China}\\
\IEEEauthorblockA{\IEEEauthorrefmark{2}School of Artificial Intelligence, Xidian University, Xi'an, China}\\
\IEEEauthorblockA{\IEEEauthorrefmark{3}School of Engineering \& Applied Science, Princeton University, Princeton, NJ}\\
E-mail: \{yongxiao, liyingyu\}@hust.edu.cn, gmshi@xidian.edu.cn, poor@princeton.edu
}
}

%
%
%

\maketitle

\begin{abstract}
This paper studies an edge intelligence-based IoT network in which a set of edge servers learn a shared model using federated learning (FL) based on the datasets uploaded from a multi-technology-supported IoT network. The data uploading performance of IoT network and the computational capacity of edge servers are entangled with each other in influencing the FL model training process. We propose a novel framework, called federated edge intelligence (FEI), that allows edge servers to evaluate the required number of data samples according to the energy cost of the IoT network as well as their local data processing capacity and only request the amount of data that is sufficient for training a satisfactory model. We evaluate the energy cost for data uploading when two widely-used IoT solutions: licensed band IoT (e.g., 5G NB-IoT) and unlicensed band IoT (e.g., Wi-Fi, ZigBee, and 5G NR-U) are available to each IoT device. We prove that the cost minimization problem of the entire IoT network is separable and can be divided into a set of subproblems, each of which can be solved by an individual edge server. We also introduce a mapping function to quantify the computational load of edge servers under different combinations of three key parameters: size of the dataset, local batch size, and number of local training passes. Finally, we adopt an Alternative Direction Method of Multipliers (ADMM)-based approach to jointly optimize energy cost of the IoT network and average computing resource utilization of edge servers. 
We prove that our proposed algorithm does not cause any data leakage  nor disclose any topological information of the IoT network. Simulation results show that our proposed framework significantly improves the resource efficiency of the IoT network and edge servers with only a limited sacrifice on the model convergence performance. 

\end{abstract}
\begin{IEEEkeywords}
6G, edge intelligence, federated learning, IoT.
\end{IEEEkeywords}
\vspace{-0.1in}

\section{Introduction}
\label{Section_Introduction}


Data is at the heart of next generation IoT networks. With the dramatic growth in demand for smart services and applications that rely on cross-domain data processing and analysis, data privacy has become one of the key issues when deploying data-driven AI-based solutions in large-scale IoT networks. Traditional cloud-based solution in which IoT devices must upload all the data into a cloud data center for a centralized process is known to be difficult to meet the increasingly stringent requirements in service responsiveness as well as data security and privacy protection.

Edge intelligence has been recently promoted as one of the key solutions to unleash the full potential of data-driven IoT networks. It inherits the decentralized nature of edge computing and emphasizes more on performing and delivering AI-functions and solutions closer to the data source, e.g., IoT devices.
%
One of the key challenges for applying edge intelligence in IoT networks is to implement distributed data processing and learning across a large number of decentralized datasets owned or managed by different user devices with privacy protection requirements. Federated learning (FL) is an emerging distributed AI framework that enables collaborative machine learning (ML) across decentralized datasets. It offers a viable solutions for data-driven learning and model training across privacy-sensitive datasets. 

Despite of its huge potential, implementing FL in edge intelligence-based IoT networks is hindered by several challenges. In particular, edge intelligence relies on a large number of low-cost edge servers to perform AI tasks and solutions. The geographical distributions and data processing capabilities of edge servers can vary significantly  causing 
highly unbalanced service demands and resource utilization. Also, the quality and size of dataset available at each edge server directly affects the performance of the model training. Unfortunately, the random nature of the wireless links connecting IoT devices and edge servers can exhibit high temporal and spatial fluctuation which may result in different numbers of data samples being uploaded from the IoT network to different edge servers. 
Currently, there is still a lack of a comprehensive framework that can coordinate and jointly optimize the data collection and transportation of the IoT network and the data processing and model training at edge servers.

In this paper, we investigate an edge intelligence-based IoT network in which an edge computing system consisting of a set of edge servers trains a shared model based on the data samples uploaded from an IoT network. 
We propose a novel framework, referred to as {\em federated edge intelligence (FEI)}, to allow jointly optimization of IoT networks and edge servers. In FEI, each edge server will 
only request a limited amount of data to be collected and uploaded from the IoT network as long as the requirement of model can be met. We consider a multi-technology-supported IoT network in which each IoT device can use licensed band IoT (LB-IoT) (e.g., NB-IoT) as well as unlicensed band IoT (UB-IoT) (e.g., Wi-Fi, 5G NR-U, Zigbee, LoRa, etc.) technologies to upload data. The main objective is to maximize the utilization of computational resources of edge servers and at the same time minimize energy cost of the IoT network. We propose a distributed algorithm that can address both problems. 
More specifically, to minimize the energy cost of IoT network, we evaluate the energy cost for uploading data samples with both LB-IoT and UB-IoT and prove that the energy cost minimization problem of the IoT network is separable and can be divided into a set of subproblems each can be solved by an individual edge server. 
For the edge computing system, we introduce a mapping function to quantify the convergence performance of model training under different combinations of three key parameters: size of the dataset, local batch size, and number of local training passes. We finally adopt an Alternative Direction Method of Multipliers (ADMM)-based approach to jointly optimize both energy cost minimization problem of the IoT network and computing resource utilization of the edge computing system. 
We prove that our proposed algorithm does not cause  data leakage of any local dataset nor reveal any network-related information such as topological information about the IoT network.  
Numerical results show that our proposed framework can significantly reduce energy cost of the IoT network and at the same time improve the resource utilization of the edge computing system with only a slight degradation of the convergence performance of FL model training. 

The remainder of this paper is organized as follows. Existing work that is relevant to this paper is reviewed in  Section \ref{Section_RelatedWork}. We introduce our system model and formulate the joint optimization problem in Section \ref{Section_SystemModel}. The FEI framework and a distributed optimization algorithm are proposed in Section \ref{Section_Architecture}. Numerical results are presented in Section \ref{Section_NumericalResult} and the paper is concluded in Section \ref{Section_Conclusion}.

\vspace{-0.1in}
\section{Related Work}
\label{Section_RelatedWork}
\noindent
{\bf Edge computing-based IoT:} Edge computing has already been recognized as one of the key components in the next generation IoT network to perform low-latency data processing and analysis. Most existing works focus on optimizing the communication resource (e.g., spectrum) allocation of IoT and computational load distribution of edge computing. For example, in \cite{Yu2018FogComputingIoT}, the authors studied the joint allocation of computing and networking resources for application provisioning in edge computing-enabled IoT networks requiring QoS guarantees. The task offloading and resource allocation problems for distributed edge computing networks have also been investigated under various time-varying scenarios\cite{Alameddine2019EdgeComputingIoTDyanmic,XY2018EHFogComputing}.


\noindent
{\bf FL for Networking Systems:}
As mentioned earlier, two major challenges that are obstructing the wide spread implementation of FL in wireless networks are heterogeneity of datasets and highly unbalanced data processing capabilities between edge servers. Many recent works have already introducing solutions to address each of these challenges. For example, solutions such as user data sharing\cite{Zhao2018FLNonIID} and user selection for model updating\cite{Wang2020FLNonIID} have been shown to be effective to reduce the adverse effect caused by the first challenge under certain scenarios. To address the second challenge, the authors in \cite{Tu2020FLFogCompNetwork} have proposed to allow discarding and offloading of data among edge servers to balance the computational load among edge servers.

In this paper, we focus on addressing both of the above challenges simultaneously by investigating the relationship between the energy cost minimization of IoT network and resource utilization optimization of edge servers. We observe that these two problems are closely related to each other. A distributed and privacy-preserving framework that can optimize both of these problems is proposed. To the best of our knowledge, this is the first work to investigate the joint optimization and coordination of  data collection and uploading of IoT and the computational resource utilization of edge servers for edge intelligence-based IoT.

\section{System Model and Problem Formulation}
\label{Section_SystemModel}
\begin{figure}
\centering
\includegraphics[width=2.6 in]{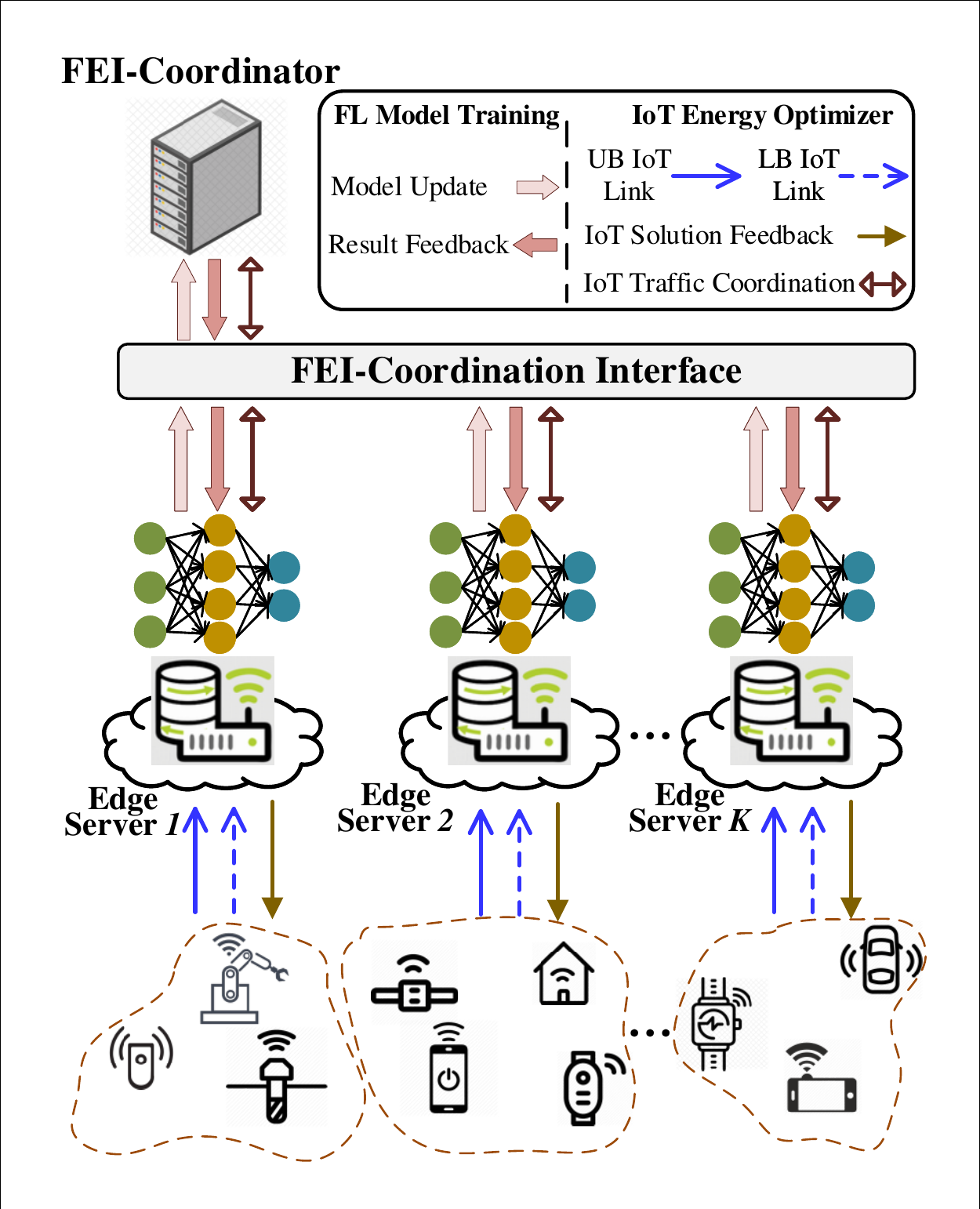}
\caption{System model for federated edge intelligence. }
\label{Figure_SystemModel}
\vspace{-0.3in}
\end{figure}

We propose  federated edge intelligence (FEI) for an IoT network to upload data samples to a set $\cal K$ of edge servers for training a share model, as shown in Figure \ref{Figure_SystemModel}. The data uploading configuration of the IoT network and the data processing and training capability of the edge computing system are entangled with each other to influence the accuracy and convergence of the model training process.

\subsection{IoT Network}
Let $\cal M$ be the set of IoT devices in the considered IoT network. Suppose the data collected by each IoT device consists of propriety information and therefore must be encrypted and uploaded to its associated edge server. Let ${\cal M}_k$ be the set of IoT devices associated with edge server $k$ for $k \in {\cal K}$ that satisfies ${\cal M}_k \subseteq {\cal M}$ and $\cup_{k\in {\cal K}}{\cal M}_k = {\cal M}$.
We assume each IoT device can adopt two types of widely adopted IoT solutions, {\em LB-IoT} and {\em UB-IoT}, to upload its data to the associated edge server.  

\noindent
{\bf LB-IoT} including 5G NB-IoT and its evolution towards B5G and 6G. In LB-IoT, each IoT device can pay for a dedicate channel in the licensed spectrum from a mobile network operator (MNO) to send its data samples. Let $\beta_m$ be the price charged by the MNO to send each data sample from IoT device $m$ to the associated edge server. We consider energy-sensitive IoT network and let $\rho_m$ be the cost for IoT device $m$ to consume each unit of energy in data uploading. This energy cost can be related to the real money spent by the IoT service operator to provide energy supplies for IoT nodes, e.g., the cost of replacing the battery or the electricity price charged by the utility company. Suppose each IoT device $m$ uploads $n''_m$ data samples in LB-IoT. The total cost for the IoT devices to upload the required data samples to edge server $k$ can then be written as
    \begin{eqnarray}
    c_k^l \left(\bn_k''\right) = \sum_{m\in {\cal M}_k} \left( \rho_m \nu^l_m \left(n''_{m}\right) + \beta_m n''_{m} \right),
    \end{eqnarray}
    where $\bn_k''=\langle n''_{m} \rangle_{m\in {\cal M}_k}$ and $\nu^l_m (n''_{m})$ is the total number of energy units required for IoT device $m$ to upload $n''_{m}$ data samples using LB-IoT.

\noindent
{\bf UB-IoT} including 5G New Radio-Unlicensed (NR-U), LoRa and ZigBee as well as their future evolution variants. In UB-IoT, all the IoT devices share the same unlicensed band for free. In this case, each IoT device only needs to pay for the energy cost for data uploading. We can write the cost for uploading ${\boldsymbol n}'_k=\langle n'_{m} \rangle_{m\in {\cal M}_k}$ data samples from set ${\cal M}_k$ of IoT devices to edge server $k$ as
    \begin{eqnarray}
    c_k^u ({\boldsymbol n}'_k) = \sum_{m\in {\cal M}_k} \rho_m {\nu}^u_m (n'_{m}),
    \end{eqnarray}
    where $\nu^u_m (n'_{m})$ is the number of energy units consumed by IoT device $m$ using UB-IoT. 
    Note that, since the unlicensed band is open to various wireless technologies, successful data delivery cannot always be guaranteed due to the possibility of collision between multiple coexisting devices.

The main objective for IoT network is to solve the following problem.  

\noindent
{\bf (P1) Energy Cost Minimization Problem}: Each edge server has limited computational resource and can only process a limited number of data samples during a given time duration. If the volume of received data exceeds the computational capacity of an edge server, IoT network can reduce the data collection and uploading rate to reduce costs. 
    Based on this observation, in our framework, edge servers will first evaluate their required size of dataset (the number of local data samples) according to the model training requirement. Each edge server will then decide the suitable data uploading configuration for its associated IoT devices including 
    the number of data samples sent from each IoT device in each available IoT solution (LB-IoT or UB-IoT).
    Motivated by the fact that, in most large IoT networks, managing the energy efficiency of the entire network is more important than minimizing the energy cost of an individual IoT device, we focus on minimizing the total energy cost of an IoT network. Suppose there is the maximum energy constraint $\epsilon$ (the maximum number of energy units) for the IoT network, i.e., we have $\nu(\bn',\bn'') \le \epsilon$ where $\nu(\bn',\bn'') = \sum_{m\in {\cal M}}\left(\nu^u_m (\bn') + \nu^l_m (\bn'')\right)$ is the total energy consumption of the IoT network.
We can then write the energy cost minimization problem of IoT network as
\begin{subequations}\label{eq_EvergyOptProblem}
\begin{align}
\mbox{\bf (P1)}\;\;\; \min_{\bn', \bn''} &\;\;\;  c(\bn', \bn'') = \sum_{k \in {\cal K}} \left(c_k^u (\bn_k') + c_k^l (\bn_k'') \right)
\\
\;\;\;\;\;\;\mbox{s.t. }& \;\; \;\;\;\;\nu(\bn',\bn'') \le \epsilon \; \mbox{ and } \; n'_m + n''_m \le {\bar n}_m,
\end{align}
\end{subequations}
where $\bn'=\langle n'_{m} \rangle_{m\in {\cal M}}$, $\bn''=\langle n''_{m} \rangle_{m\in {\cal M}}$, and ${\bar n}_m$ is the maximum number of data samples that can be collected by each IoT device $m$.

\subsection{Edge Computing System}
Recent results\cite{McMahan2017FLfirstpaper,Bonawitz2019FLMassiveDeploy} as well as our own experiments have suggested that for a given type of dataset (e.g., MNIST, CIFAR, etc.), the computational load in each edge server $k$ to train an FL model is closely related to three key parameters: (1) size of the local dataset, denoted as $n_k$, (2) the number of local training passes in each coordination round, denoted as $e_k$, and (3) the minibatch size of each local training pass, denoted as $b_k$. Motivated by this observation, we introduce an empirical-based mapping function $\mu_{k}$ to quantify the relationship between computational load and combinations of parameters $e_k$, $b_k$, and $n_k$ under a given model convergence requirement, i.e., there exists a computational load function for each edge server $k$ given by ${\tilde \mu_k } = \mu_{k}\left( b_k, e_k, n_k | t, r \right)$ where $t$ denotes the type of dataset and $r$ is the convergence requirement of the model training (minimum required model accuracy under a given number of rounds of model updating). We will give a more detailed discussion in Section \ref{Section_Architecture}.

We investigate the following problem for edge computing system.

\noindent
{\bf (P2) Resource Utilization Maximization of Edge Computing System}: 
Let ${\bar \mu}_k$ be the maximum computational load that can be supported by the local computational resources of edge server $k$, i.e., we have ${\tilde \mu_k } < {\bar \mu}_k$. Suppose the number of data samples arrived at edge server $k$ is given by $n_k = \sum_{m\in {\cal M}_k} \left(n'_{m}+n''_m\right)$. We can then define the average resource utilization maximization problem for an edge computing system as
\begin{subequations}\label{eq_EdgeCompuOptProblem}
  \begin{align}
\mbox{\bf (P2)}\;\;\; \min_{\bb, \be, \bn}  & \;\;  \mu \left( {\bn} \right) = {1\over K}\sum_{k\in {\cal K}} \left( {\bar \mu}_{k} - {\tilde \mu}_{k} \right) \\
\;\;\;\;\;\;\mbox{s.t. }  & n_k < {\tilde n}_{k} \; \mbox{ and } \; {\tilde \mu}_{k} \le {\bar \mu}_{k}, \forall k \in {\cal K},
  \end{align}
\end{subequations}
where ${ \bb}=\langle b_{k} \rangle_{k\in {\cal K}}$, ${ \be}=\langle e_{k} \rangle_{k\in {\cal K}}$, ${\bn}=\langle n_{k} \rangle_{k\in {\cal K}}$, 
and ${\tilde n}_{k}$ is the maximum number of data samples that can be received and processed by edge server $k$. 

\vspace{-0.6em}
\subsection{Joint Optimization of Resource Efficiency}
It can be observed that problems (P1) and (P2) are closely related to each other. To investigate the joint  optimization of these problems, we introduce a weighing factor $\gamma$ specifying the relative weights between energy cost of the IoT network and computational resource utilization of the edge computing system. In this paper, we focus on solving the following problem:
\vspace{-0.6em}
\begin{subequations}\label{eq_JointOptProblem}
\begin{align}
\mbox{\bf (P3)} \;\;\; \min_{\bn', \bn''}  &\;\;  \mu \left( {\bn} \right) + \gamma c(\bn', \bn'') \label{eq_JointOptProblemObj}\\
\mbox{s.t. } & \;\; \nu(\bn',\bn'') \leq \epsilon \label{eq_JointOptProblemCb}\\
\;\;\;\;\;\;\;\;&\;\; n'_m+n''_m\le {\bar n}_m, \;\forall m\in {\cal M}\label{eq_JointOptProblemCc}\\
\;\;\;\;\;\;\;\;&\;\;  n_k < {\tilde n}_{k} \; \mbox{ and } \; {\tilde \mu}_{k} \le {\bar \mu}_{k}, \forall k \in {\cal K}\label{eq_JointOptProblemCd},
\end{align}
\end{subequations}

Solving problem (\ref{eq_JointOptProblem}) requires the full knowledge of both the IoT network and edge computing system which may not always possible, especially in a distributed networking system where the data information of each edge server cannot be revealed to others. In the rest of this paper, we introduce a novel distributed framework 
that can jointly coordinate edge servers to train a FL model across edge servers and at the same time minimize the energy cost of IoT network by optimizing the data uploading traffic distribution between UB-IoT and LB-IoT across all the IoT devices. Our framework will not disclose any data-relevant  information.


\vspace{-0.1in}
\section{FEI Framework and Resource Efficiency Optimization}
\label{Section_Architecture}

\begin{figure}
\centering
\includegraphics[width=3.3 in]{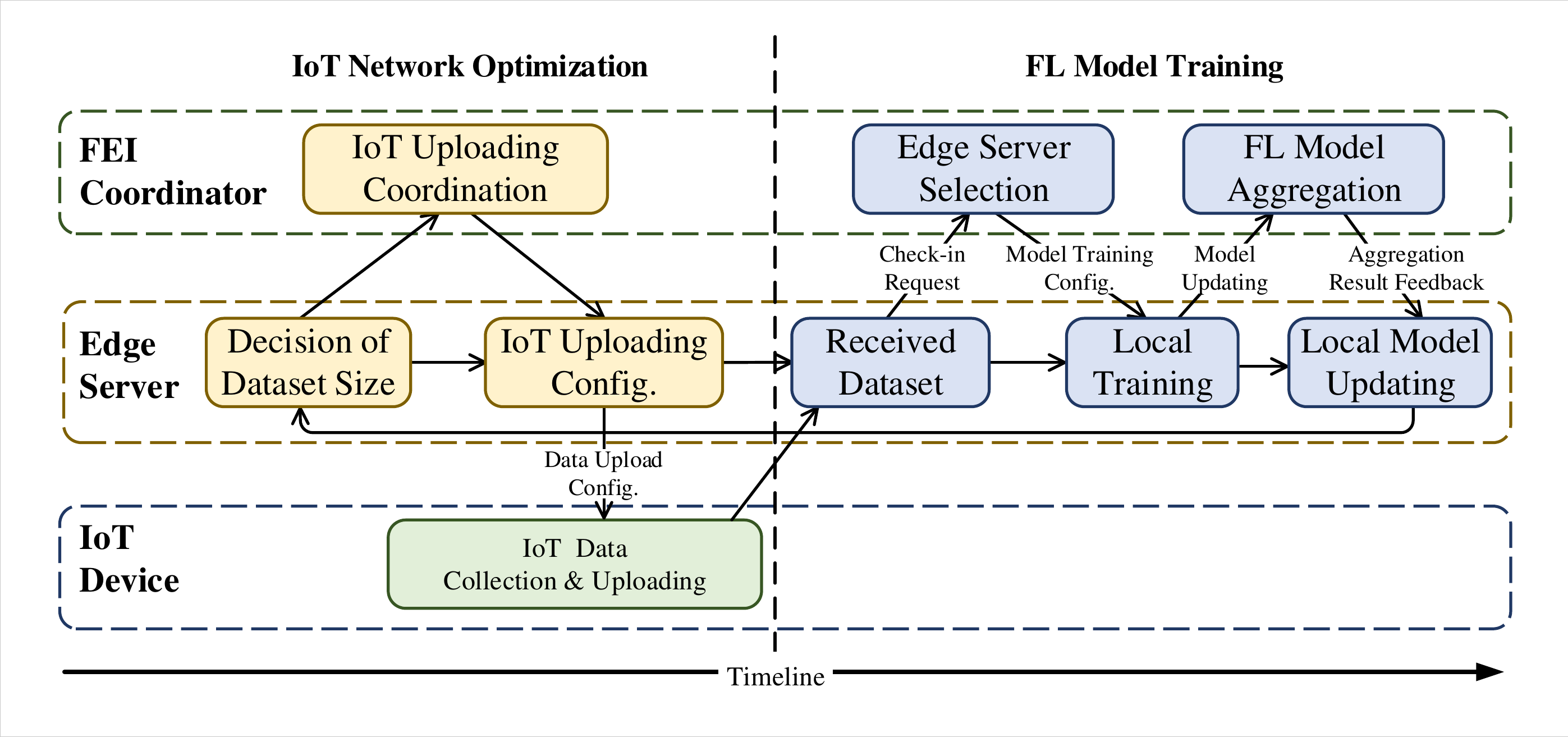}
\caption{Proposed framework and procedure of the FEI. }
\label{Figure_procedure}
\vspace{-0.1in}
\end{figure}

\subsection{Framework Overview}
The detailed procedure of the FEI framework is illustrated in Figure \ref{Figure_procedure}. 
In FEI, each edge server will first evaluate the required size of dataset to be uploaded from the IoT network. We propose a distributed coordination approach based on distributed ADMM for edge servers to coordinate via FEI coordinator and jointly decide the size of data samples to be uploaded from each IoT device. In this approach, each edge server will initially 
randomly generate a number of data samples to be uploaded from the IoT network. The generated number needs to satisfy the data collection constraint of IoT devices as well as its data processing limit. Edge servers will then submit their randomly selected numbers to the FEI coordinator. 
FEI coordinator will calculate an intermediate result based on the received numbers and send the result back to the edge servers. Each edge server will use this intermediate result to calculate an updated solution about the number of data samples to be requested from the IoT network. The above process will be repeated and 
the result will converge to the global optimal solution of 
dataset sizes to be uploaded from IoT devices with each IoT solution (UB-IoT or LB-IoT). We will give a more detailed discussion 
in Section \ref{Subsection_DistributedAlgorithm}).  

After receiving the required number of data samples, each edge server will then proceed to the FL model training procedure. In this procedure, edge servers will first decide whether to register with the FEI coordinator to join the next round of model training. FEI coordinator will select a subset of edge servers to participate in the training and will send the detailed configuration information, such as data structure, sharing state, and model parameters, to each participating edge server. Each participating edge server will perform local training computations based on the received configuration information as well as its local dataset. After the local training, edge servers will send their updates to the FEI coordinator for model aggregation. Once FEI coordinator receives enough updates, it will send the aggregation result back to participating edge servers. The edge servers will update their respective models using the received results. The above process will be repeated in the next round with a newly selected subset of edge servers until the trained model converges nor the stopping criteria have been met. 

As will be proved in Section \ref{Subsection_DistributedAlgorithm}, our proposed framework allows distributed coordination and collaborative model training between edge servers without requiring any exchange of datasets between edge servers or disclosing any information related to the IoT network including the topological information, spectrum usage, etc.


%
%
%

\subsubsection{Computational Load and Resource Demand Mapping} We follow a commonly adopted setting\cite{Bonawitz2019FLMassiveDeploy,McMahan2017FLfirstpaper} and consider an FL process with synchronized model updating in which all the edge servers will use the data collected during the same period of time with duration $\tau$ to train a globally shared machine learning model.
We assume the data samples collected by each individual IoT device at different time within each considered duration are equally important for model training. Let $\bx_k$ be the dataset collected by edge server $k$ during $\tau$. Suppose for a given $\tau$, the number of model updating rounds among edge servers can be considered as fixed and the model convergence performance can then be specified by the model accuracy that can be achieved under a fixed number of  model updating rounds. We use $r$ to denote the model convergence requirement, i.e., $r$ is the minimum model accuracy required for training within a fixed number of updating rounds.
As mentioned earlier, the computational load in each edge server $k$ to train an FL model with a certain requirement $r$ is closely related to three key parameters, including size of the local dataset $n_k$, the number of local training passes $e_k$ in each coordination round, and the minibatch size $b_k$ of each local update.

It has been shown in \cite{McMahan2017FLfirstpaper} that FL can only converge to an unbiased solution if all the edge servers have equal probability to participate in the model training. We therefore assume 
the probability for each edge server to participate each round of model update can be regarded as a constant, denoted as $C$. The value of $C$ is closely related to the communication cost among edge servers to train the model. How to optimize the value of $C$ is out of the scope of this paper and will be our future work. Suppose the size of dataset $\bx_k$ (number of samples) is $n_k$. The number of local updates  performed by edge server $k$ for each round is given by ${{e_k n_k} \over b_k}$, e.g., if each edge server uses all the received dataset for model training per pass, we have $n_k = b_k$ and, in this case, edge server $k$ will perform $e_k$ local passes for each coordination round. 
It is known that the values of $e_k$ and $b_k$ affect not only  the computational load at edge server $k$, but also the convergence performance of the entire model training process. In particular, increasing $e_k$ or decreasing $b_k$ will generally result in more computational load at edge server $k$. The impact of $e_k$ and $b_k$ on the convergence of FL model training is more complicated and will also depend on the type of the dataset. For example, for some document recognition dataset consisting of samples with highly correlated features, such as MNIST\cite{deng2012mnist}, reducing $b_k$ and increasing $e_k$ will accelerate the convergence performance. However, for some other types of datasets with more complex features, e.g., Shakespeare LSTM dataset\cite{McMahan2017FLfirstpaper}, increasing $e_k$ may result in diverged result due to the over-optimization. In this case, it is better to adopt a larger $b_k$ and a smaller $e_k$.
We present the average computational time of Federated Averaging algorithm proposed in \cite{McMahan2017FLfirstpaper} for each round of model updating using MNIST dataset under the same computation environment and setup in Table \ref{Table_ComputCompare}.

To simulate the data generated by IoT network during multiple periods, we assume each edge server receives the different subsets of samples for training different FL models at different time periods. We can observe that for a given size of the dataset, different combinations of parameters result in different convergence rate and computational load, e.g., a small $b_k$ and a large $e_k$ achieve the fastest convergence speed with the highest computational load requirement.
In this paper, we focus on a finite set $\cal T$ of dataset types  for IoT networks. We assume data samples from the same type of dataset can be considered as independent and identically distributed and contribute equal to the model training. We also assume, for a given type of dataset, there exists a one-to-one mapping from each possible combination of $b_k$, $e_k$, and $n_k$ to the computational load ${\tilde \mu_k}$ at each edge server $k$, i.e., we have ${\tilde \mu_k } = \mu_{k}\left( b_k, e_k, n_k | t, r \right)$ for ${\tilde \mu_k } < {\bar \mu}_k$, where ${\bar \mu}_k$ is the maximum computation power that can be offered by edge server $k$. We also assume for each given type of dataset $t\in {\cal T}$ with a fixed size $n_k$, there exists an optimal pair of $b_k$ and $e_k$ that offers the fastest convergence performance for FL model training, denoted as $\langle b^*_k, e^*_k \rangle =  {\cal F} \left(  n_k, t, r \right)$ for all $k \in {\cal K}$.

\begin{table}[!htbp]
 \centering
 \caption{Computation Time Comparison Under Different Parameters}
\begin{tabular}{cccccc}
 \hline
 \hline
 $n_k$ & $e_k$ & $b_k$ &\makecell[c]{Accuracy\\ (in 50 rounds)}&\makecell[c]{Accuracy\\ (in 200 rounds)}&\makecell[c]{Time per round\\ (sec)} \\
 \hline
    100 &5&10&97.12\%&98.39\%&0.1216\\
    \hline
    100 &5&50&95.34\%&97.72\%&0.0655\\
    \hline
    {\bf 100} &{\bf 20}& {\bf 10}& {\bf 97.22}\%& {\bf 98.47} \%& {\bf 0.4772}\\
    \hline
    100&20&50&95.98\%&97.81\%&0.2610\\
    \hline
     200 & 5&10&98.00\%&98.73\%&0.2386\\
    \hline
    200&5&20&97.74\%&98.67\%&0.1693\\
    \hline
    200 &5&50&96.98\%&98.36\%&0.1299\\
    \hline
    {\bf 200} &{\bf 20}& {\bf 10}& {\bf 98.04}\%& {\bf 98.85} \%& {\bf 0.9571}\\
    \hline
    200&20&20&97.88\%&98.74\%&0.6784\\
    \hline
    200&20&50&97.60\%&98.44\%&0.5200\\
    \hline
    400&5&10&98.58\%&99.01\%&0.4775\\
    \hline
    400&5&20&98.27\%&98.89\%&0.3383\\
    \hline
    400&5&50&97.87\%&98.78\%&0.2597\\
    \hline
    {\bf 400}&{\bf 20}&{\bf 10}&{\bf 98.61}\%&{\bf 99.16}\%&{\bf 1.9128}\\
    \hline
    400&20&20&98.39\%&98.97\%&1.3539\\
    \hline
    400&20&50&98.22\%&98.83\%&1.0409\\
    \hline
\end{tabular} \label{Table_ComputCompare}
\end{table}

\subsubsection{Energy Consumption of IoT Network}
Due to the different spectrum sharing regulations, LB-IoT and UB-IoT have different characteristics in terms of spectrum access, energy consumption, as well as data uploading performance. In this paper, we adopt a commonly used formulation to model the energy consumption of IoT data uploading.

\noindent
{\bf UB-IoT:} Since unlicensed band is shared by all the devices without coordination, a certain channel access mechanism must be imposed to avoid collision among spectrum sharing devices. For example, Carrier Sense Multiple Access with Collision Avoidance (CSMA/CA) mechanism has been adopted in 2.4 GHz and 5 GHz unlicensed bands for supporting UB-IoT such as Wi-Fi, ZigBee, Thread, Z-Wave, and Wi-SUN\cite{Bianchi1996CSMA}. In this mechanism, each IoT device  must  first  sense  the  vacancy  of  the  channel and can only start transmission if the channel is sensed idle. If the channel is busy, it will go through a random backoff procedure and will only be able to send data if all the procedure is finished and the channel is clear. According to \cite{XY2018UnlicencedNetSlice,Hirzallah2018DySPAN}, the probability of channel access is closely related to the topology (relative location) of all the spectrum sharing devices as well as the channel contending parameters adopted by each IoT device. If the topology and parameters of IoT devices are fixed, the probability for each IoT device to successfully occupy the channel to transmit data can be considered as fixed. Similarly, in the duty-cycle-based mechanism adopted in 433 MHz and 800 MHz bands, IoT devices do not have to sense the channel but will have to send its data packets according to a fixed duty-cycle specifying the maximum fraction of time that can be occupied by each device to access the channel, e.g., depending on the frequency bands, the maximum duty-cycle is generally between 0.1\% and 10\%. 

    Let $P^u_m$ be the probability for IoT device $m$ to successfully occupy unlicensed band to send data. If IoT node $m$ needs to upload $n'_{m}$ bits of data to edge server $k$ within time duration $\tau$, it will need to send the data at a rate of ${n'_{m}\over \tau} =  P^u_m B^u \log_2 \left( 1 + h^u_{m} {\tilde{\nu}_m^{(1)} \over \sigma^u_{m}} \right)$ where $B^u$ is the bandwidth of unlicensed channel, $h^u_{m}$ is the channel gain between IoT node $m$ and the associated UB-IoT access point (AP)\footnote{In this paper, we assume the major bottleneck for data transmission between IoT devices and edge servers is the wireless link connecting the IoT device and AP.}, and $\sigma^u_{m}$ is the received noise level at the AP. The energy consumed by IoT device $m$ for sending $n'_{m}$ data bits via a UB link can then be written as
\begin{eqnarray}\label{miu1}
\nu^u_m (n'_{m}) =  {\sigma^u_{m} \over h^u_{m}} \left( 2^{n'_{m} \over \tau B^u P^u_m}-1 \right).
\end{eqnarray}



\noindent
{\bf LB-IoT:} Each IoT device can purchase IoT connectivity service from an MNO to upload its data through the licensed band. In this case, a dedicate frequency band will be allocated to each IoT device for data transmission. Let $\beta_m$ be the price paid by IoT device $m$ for sending each bit of data via LB-IoT network infrastructure, i.e., the price paid by IoT $m$ for sending $n''_{m}$ bits of data is given by $\beta_m n''_{m}$. Similar to the previous case, we can write the energy consumed by IoT node $m$ to upload $n''_{m}$ bits of data to edge server $k$ as
\begin{equation}\label{miu3}
\nu_m^{l}(n''_{m})= {\sigma^l_{m} \over h^l_{m}}\left( 2^{n''_{m} \over \tau B^l}-1 \right),
\end{equation}
where $B^l$ is the bandwidth of each licensed channel, $h^l_{m}$ is the channel gain between IoT node $m$ and the associated LB-IoT base station (BS), and $\sigma^l_{m}$ is the received noise level at the BS. 

\vspace{-0.1in}
\subsection{Resource Efficiency Optimization Algorithm Design}
\label{Subsection_DistributedAlgorithm}
We propose a joint optimization algorithm to address Problem (P3) under the above framework based on an FEI coordinator. In this algorithm, we divide the global traffic allocation problem into $K$ subproblems each of which can be solved by an edge server, and the intermediate calculation results will be sent to the FEI coordinator that can perform coordination among all edge servers. 

To make our description clearer, we define $\bn_k = \langle n_m', n_m'' \rangle_{m \in {\cal M}_k}$ to include all the data sample optimization variables of the IoT nodes associated to each edge server, i.e., $\bn_k' =A_k \bn_k$ and $ \bn_k'' =B_k \bn_k$, where $A_k$ and $B_k$ are matrices in $R^{M_k \times 2M_k}$ with elements
\begin{equation} \label{eq_AB}
A_k(p,q) =
  \begin{cases}
    1,       & q = 2p-1,    \\
    0, & \mbox{otherwise},
  \end{cases} \mbox{and} \;
 B_k(p,q) =
  \begin{cases}
    1,       & q = 2p,    \\
    0, & \mbox{otherwise},
  \end{cases}
\end{equation}
for all $p = 1,2,...,M_k$ and $ q = 1,2,...,2M_k$, where $M_k$ is the number of IoT devices in ${\cal M}_k$.

It can be proved that by substituting (\ref{miu1})-(\ref{eq_AB}) into (\ref{eq_JointOptProblem}), it becomes a convex optimization problem in which the objective function (\ref{eq_JointOptProblemObj}) and constraints (\ref{eq_JointOptProblemCc})-(\ref{eq_JointOptProblemCd}) are completely separated, while constraint (\ref{eq_JointOptProblemCb}) is coupled across all the edge servers.
In order to distributively solve problem (\ref{eq_JointOptProblem}) distributively  without requiring each server to disclose any data or network information, we first introduce the following indicator functions to incorporate the inequality constraints (\ref{eq_JointOptProblemCb})-(\ref{eq_JointOptProblemCd})  into the objective function  (\ref{eq_JointOptProblemObj}):
\begin{equation}
I_{\cal D}(\bn) =
  \begin{cases}
    0,       & \bn \in {\cal D},    \\
    \infty, & \bn \notin {\cal D},
  \end{cases} \;\;
  I_{{\cal D}_k}(\bn_k) =
  \begin{cases}
    0,       & \bn_k \in {\cal D}_k,    \\
    \infty, & \bn_k \notin {\cal D}_k,
  \end{cases}
\end{equation}
where $\bn = (\bn_1; \bn_2; ... ; \bn_K)$, ${\cal D} = \{\bn|\sum_{k \in {\cal K}}  \nu_k(\bn_k) \leq \epsilon\}$, and $ {\cal D}_k = \{\bn_k| \bn_k \; \mbox{satisfies (\ref{eq_JointOptProblemCc}-\ref{eq_JointOptProblemCd})}  \}$.
The objective function for each subproblem to be solved by an edge server can be rewritten as:
\begin{equation}\label{eq_JointOptProblemSepfk}
 f_k(\bn_k) =   {1\over K} \left( {\bar \mu}_{k} - {\tilde \mu}_{k} \right) + c_k^u (A_k\bn_k) + c_k^l (B_k\bn_k).
\end{equation}

We can then reformulate problem (\ref{eq_JointOptProblem}) in a standard ADMM form as
\begin{subequations}\label{eq_JointOptADMM}
\begin{align}
\min_{\bn, \bw}  &\;\; \sum_{k \in {\cal K}} \left( f_k(\bn_k) +  I_{{\cal D}_k}(\bn_k) \right) + I_{\cal D}(\bw), \\
\mbox{s.t. } & \;\; \bn - \bw = {\boldsymbol 0},
\end{align}
\end{subequations}
where  $\bw = \langle \bw_1; \bw_2; ... ; \bw_K \rangle$ is the introduced auxiliary variable. We can therefore adopt the following updates to solve problem (\ref{eq_JointOptADMM}) iteratively:
\begin{subequations}
\begin{align}
\{\bn_k^{i+1}\}_{k \in {\cal K}} & = \mathrm{\mathop {argmin}\limits_{\bn_k}}  \left( f_k(\bn_k) +  I_{{\cal D}_k}(\bn_k) \right. \nonumber \\
&\;\;\;\;\;\;\;\;\;\;\;\;\;\;\;\;\;\;\; \left. + \frac{\eta }{2}\| \bn_k- \bw_k^{i} + \btheta_k^{i}\|_2^2\right)
  \label{eq_ADMMn},\\
\bw^{i + 1} &= \mathrm{\mathop {argmin}\limits_{\bw}} \left( I_{\cal D}(\bw) +  \frac{\eta }{2}\| \bn^{i+1}- \bw + \btheta^{i}\|_2^2 \right)
\label{eq_ADMMw},\\
{\btheta^{i + 1}} &= \btheta^{i}+\bn^{i+1}-\bw^{i+1},\label{eq_ADMMtheta}
\end{align}
\end{subequations}
where $i$ is the iteration index,  $\btheta = (\btheta_1; \btheta_2; ... ; \btheta_K)$ is the dual variable, and $\eta$ is the augmented Lagrangian parameter. We summarize the details in Algorithm~\ref{Algorithm} where the stopping criteria follow the same line as that in \cite{ADMM}.
\begin{algorithm}[!t]
  \caption{Resource-Efficiency Optimization Algorithm}\label{Algorithm}
  \begin{algorithmic}
    \STATE \textit{Initialization}: $\bn^0 \in R^{2M\times 1} $, $\eta>0$;
    \FOR {$i=0,1,2,...$}
    \STATE 1. Each edge server $k$ executes the following steps using its local information:
    \STATE \quad 1) Calculate $\bn_k^{i+1}$ according to (\ref{eq_ADMMn});
    \STATE \quad 2) Broadcast $\bn_m'$ and $\bn_m'$ to each IoT device $m \in {\cal M}_k$, so it can allocate its traffic distribution accordingly;
    \STATE \quad 3) {\bf if} Stopping criteria have been met  {\bf then}
    \STATE \quad \quad  \quad \quad Break;
    \STATE \quad \;\; {\bf else}
     \STATE \quad \quad \quad \quad Report $\bn_k^{i+1}$ to the FEI coordinator;
    \STATE \quad \;\; {\bf end if}
    \STATE 2. When the FEI Coordinator receives all the updates of $\bn_k^{i+1}$, it will execute the following steps:
    \STATE \quad 1) Update the $\bw^{i+1}$ and $\btheta^{i+1}$ according to (\ref{eq_ADMMw}) and  (\ref{eq_ADMMtheta});
    \STATE \quad 2) Feedback $\bw_k^{i+1}$ and $\btheta_k^{i+1}$ to each edge server $k$.
    \ENDFOR
  \end{algorithmic}
\end{algorithm}

\begin{theorem}\label{theorem}
Algorithm~\ref{Algorithm} converges to the optimal solution of Problem (P3) without requiring any exchange of local datasets among edge servers nor disclosing any IoT network topological information.
\end{theorem}
\begin{IEEEproof}
We briefly describe our proof as follows:
\begin{enumerate}
  \item The FEI coordinator only receives the intermediate traffic allocation results from edge servers which cannot be used to reveal any content information of their local datasets.
  \item Each edge server only receives  subvectors  $\bw_k^{i+1}$ and $\btheta_k^{i+1}$ from the FEI coordinator. It cannot derive any information about other edge servers and IoT devices.
  \item Each IoT device only receives traffic allocation results $\bn_m'$ and $\bn_m'$ from its associated edge server which do not contain any information about other IoT devices or edge servers.
\end{enumerate}
\end{IEEEproof}

\vspace{-0.2in}
\section{Numerical Result}
\label{Section_NumericalResult}


\begin{figure}
\centering
\begin{minipage}[t]{0.4\linewidth}
\centering
\includegraphics[width=1.4 in]{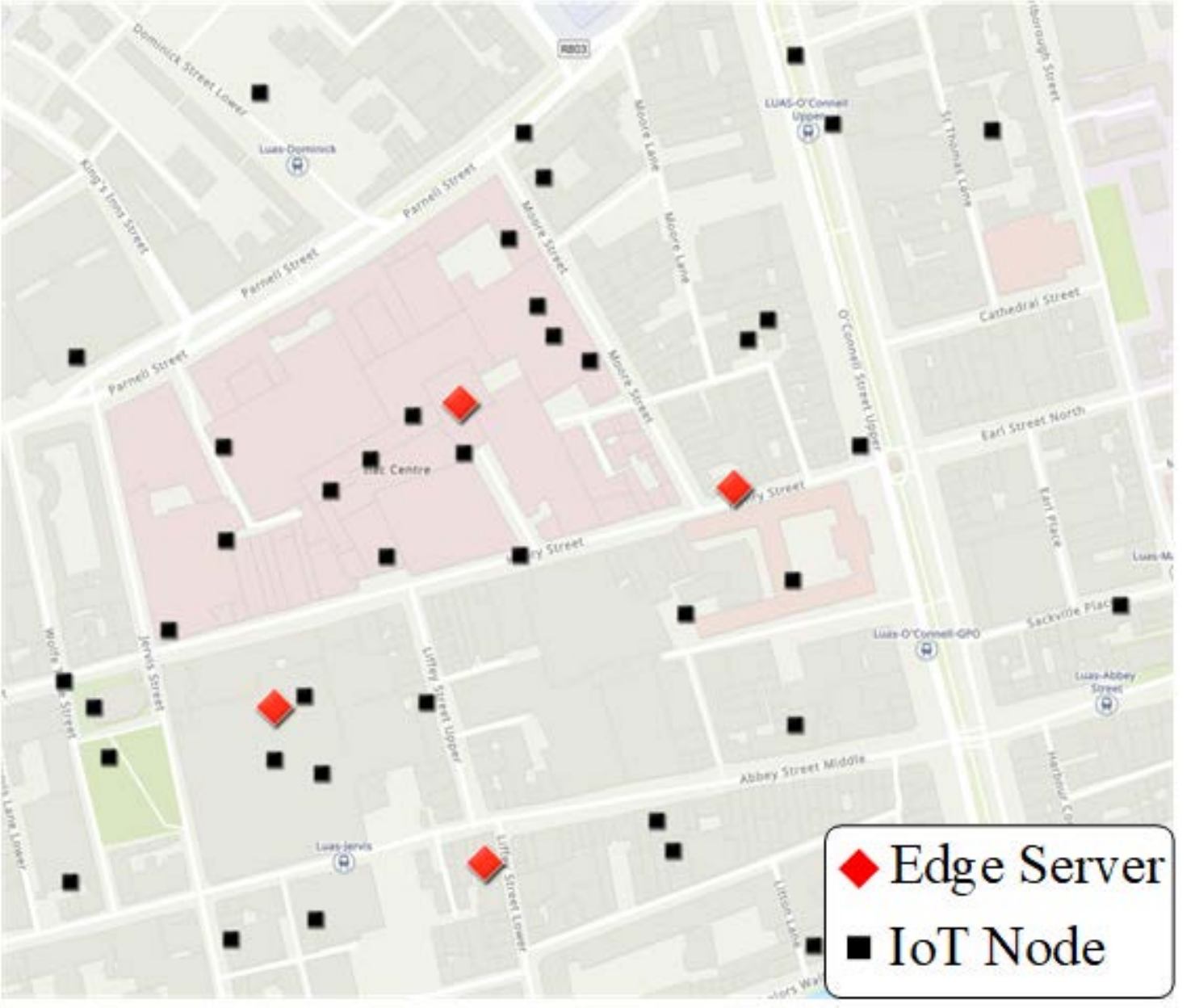}
\vspace{-0.15in}
\caption{Simulation setup.}
\label{Figure_SimSetup}
\end{minipage}
\begin{minipage}[t]{0.55\linewidth}
\centering
\includegraphics[width=1.6 in]{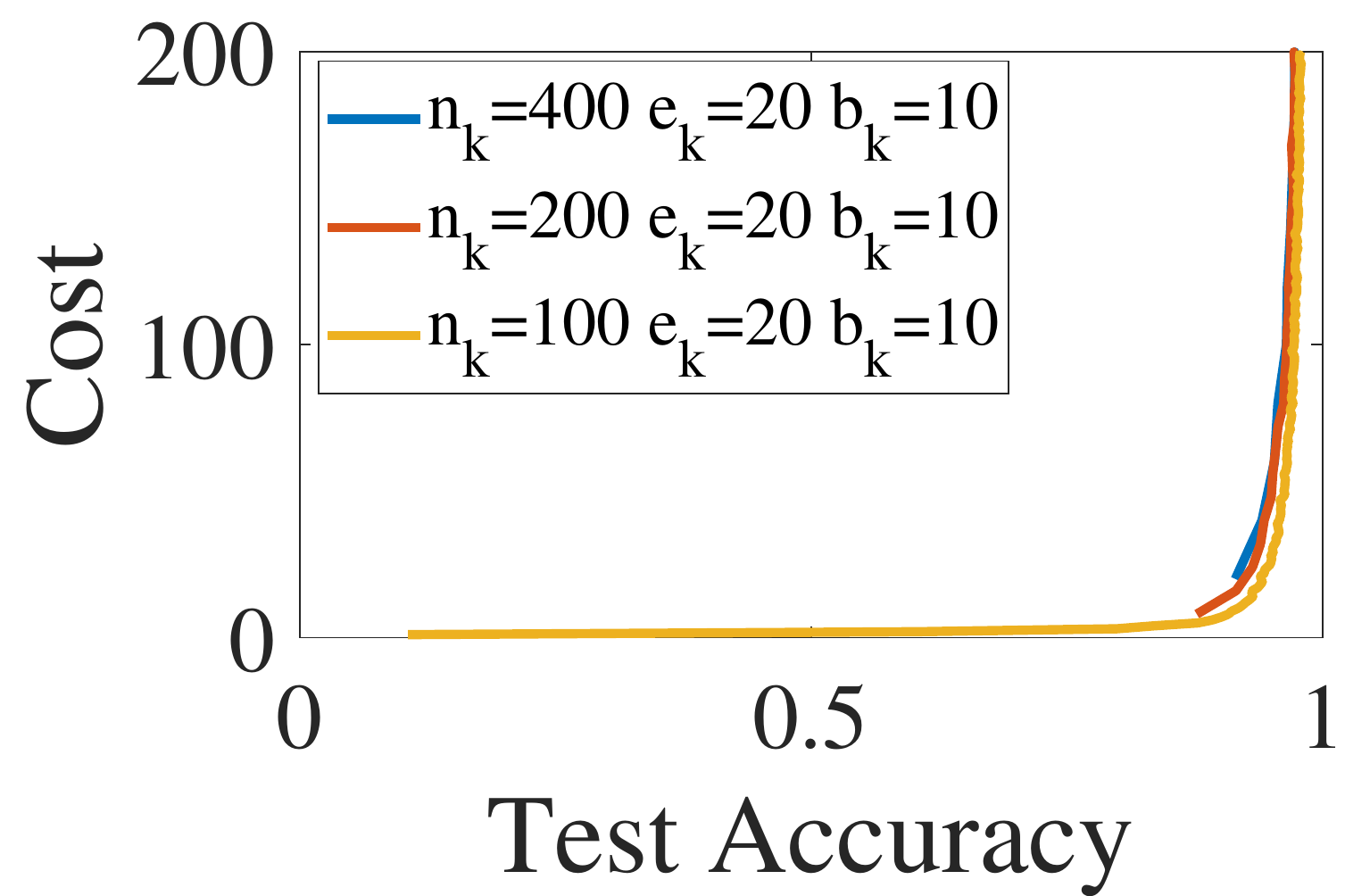}
\caption{Cost of FEI under different requirements of model accuracy.}
\label{Figure_CostVSAcc}
\end{minipage}
\begin{minipage}[t]{0.45\linewidth}
\centering
\includegraphics[width=1.6 in]{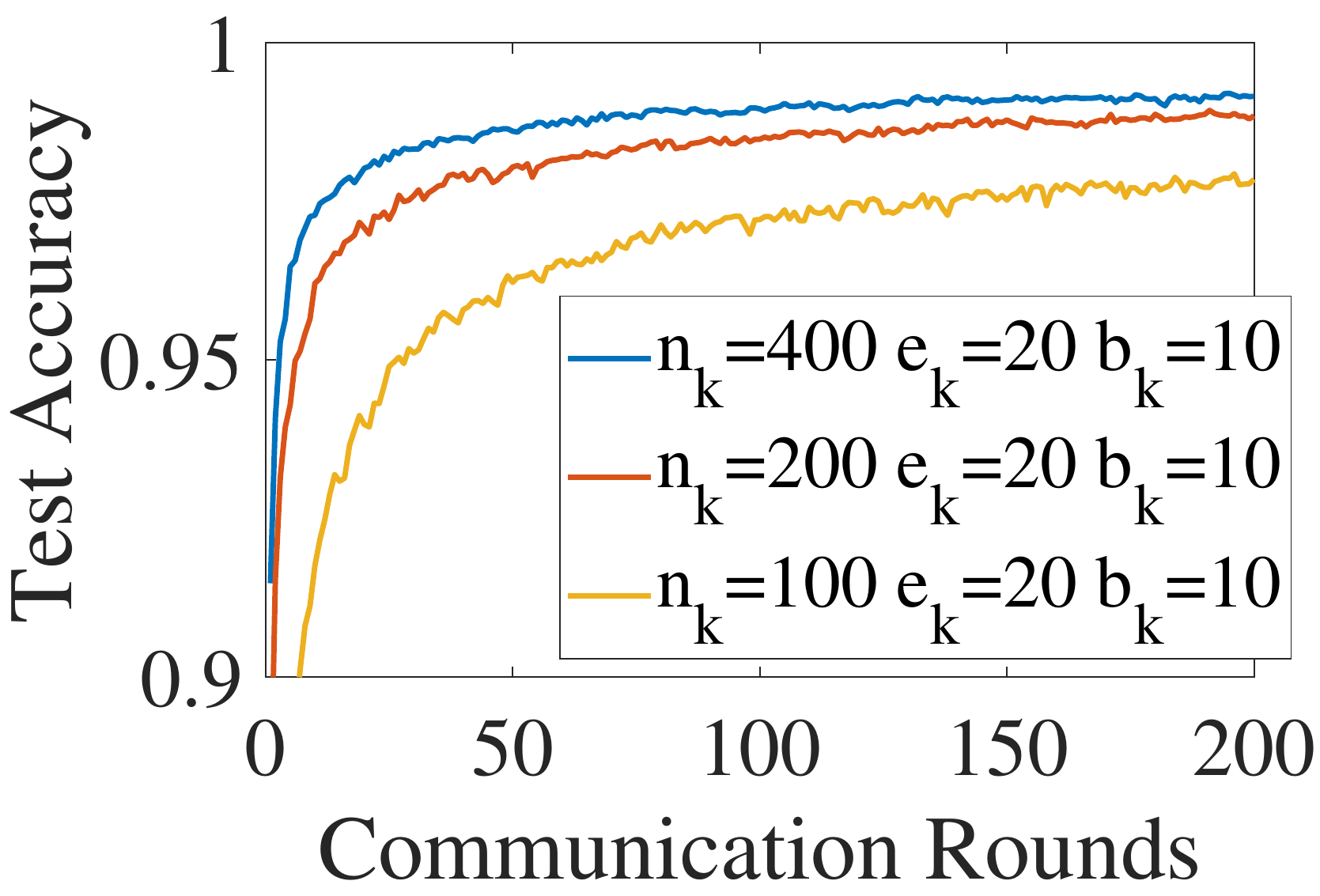}
\vspace{-0.2in}
\caption{Convergence performance under different model training parameters.}
\label{Figure_IoTOptAccVSRounds}
\end{minipage}
\begin{minipage}[t]{0.45\linewidth}
\centering
\includegraphics[width=1.6 in]{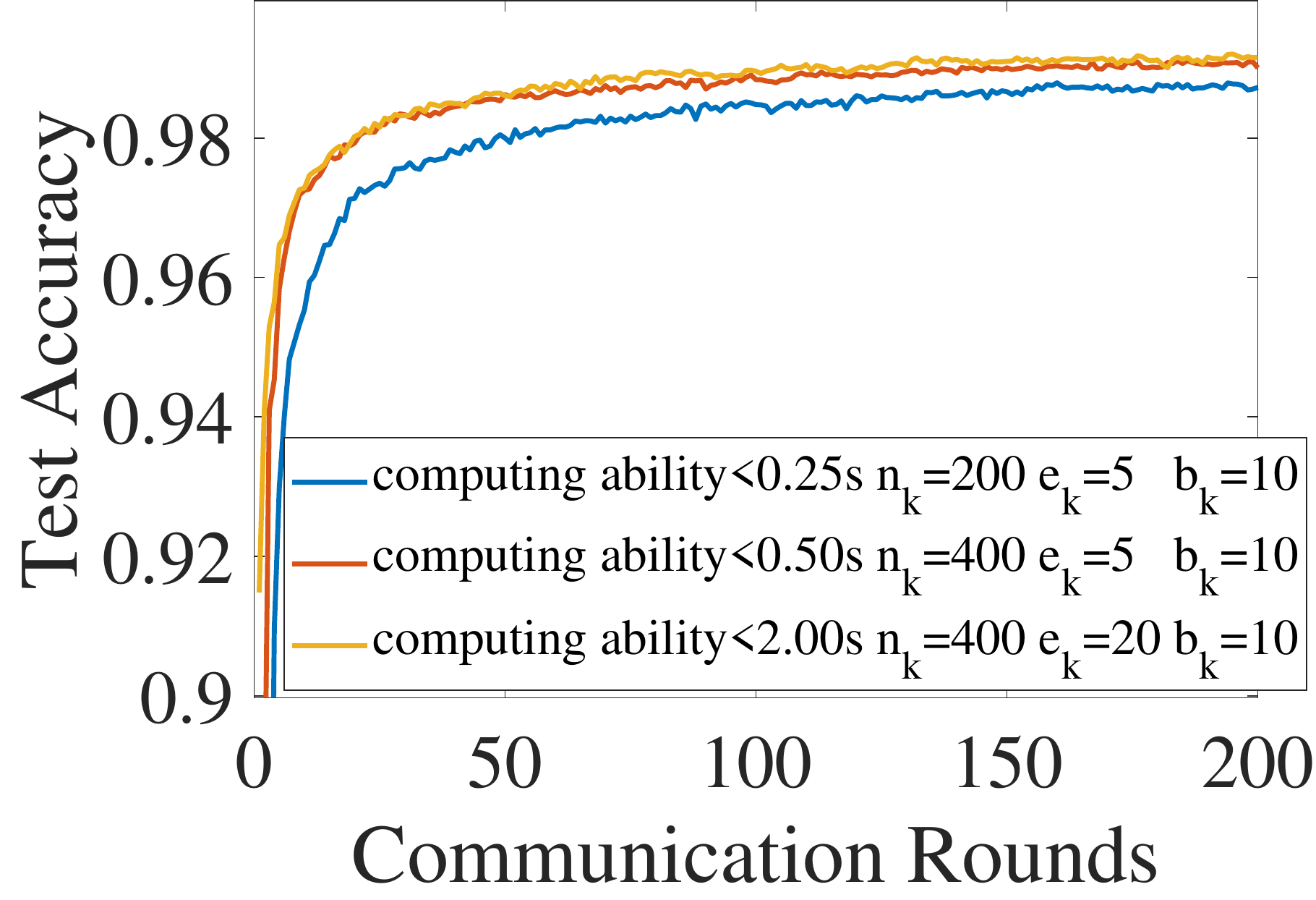}
\vspace{-0.2in}
\caption{Convergence performance under different computational capacities of edge server.}
\label{Figure_AccVSEdgeCapacity}
\end{minipage}
\vspace{-0.2in}
\end{figure}

To evaluate the performance of our proposed framework, we simulate an IoT network that can upload data samples to 4 edge servers using the real locations of Starbuck's coffee shops in the city center of Dublin city as shown in Figure \ref{Figure_SimSetup}. We assume that each edge server can receive data from 10 IoT devices. We use the data samples from the EMNIST dataset to simulate the data uploaded from each IoT device to an edge server. We equally divide 341873 training examples and 40832 test examples among 4 groups of IoT devices and in each round of data uploading, each IoT device randomly selects data samples from its local training  example pool to upload to the associated edge server. The average sample size of EMNIST is 0.67 KB. We assume each IoT device can use Zigbee or NB-IoT to submit data samples with transmit powers $63$ mW and $100$ mW, respectively. We simulate each edge server with a TITAN X GPU running on Ubuntu 16.04. The training process among 4 edge servers is simulated using Federated-Tensorflow framework 0.16.1. Each edge server trains a four-layer CNN model locally. 

We first evaluate the impact of the data uploading performance of IoT network on the convergence rate of the FL model training. In Figure \ref{Figure_CostVSAcc}, we consider a fixed computational capacity of each edge server (computational time for each round of training is less than 0.25s, 0.5s, and 2s, respectively) and compare the total cost under different model accuracies  when the IoT network can use only LB-IoT, only UB-IoT, and optimal data uploading configuration with both IoT technologies. We can observe that the optimal data uploading scheme significantly reduces the total cost. The LB-IoT however results in the highest cost among all the IoT solutions. We then present the model accuracy under different communication rounds in the same setup in Figure \ref{Figure_IoTOptAccVSRounds}. In this case, we can observe that LB-IoT achieves the best convergence performance among all three solutions. This is because LB-IoT can always upload the required amount of data samples to edge servers during every round of training. The optimal uploading scheme however tries to balance the update uploading performance and the cost and therefore results in slightly reduced convergence performance. We can also observe that, with only a small degradation of convergence performance of the training process, the cost of the IoT network can be significantly reduced. In other words, our proposed optimal uploading scheme is more suitable for cost-sensitive IoT network that is not very sensitive to the accuracy or convergence of the model.

To evaluate the impact of the computational capacity of edge servers on the model training performance, we compare the convergence performance under different upper limits of the computational load (maximum computational time per round of training) in Figure \ref{Figure_AccVSEdgeCapacity}. We can observe that  the increasing speed of the model training performance in terms of convergence speed and accuracy improvement reduces with the computational capacity. Therefore, for IoT networks with limited model accuracy requirements, deploying low-cost edge servers with limited computational capacity may be sufficient for most FL model training tasks.

\vspace{-0.1in}
\section{Conclusion}
\label{Section_Conclusion}

This paper has introduced the FEI framework to jointly optimize the data collection and transportation of an IoT network and resource utilization of an edge computing system. FEI allows edge servers to first evaluate the required number of data samples according to its local data processing capacity and requirements of the FL model and will request a sufficient amount of data with minimized energy cost for the IoT network. The energy cost of the IoT network has been evaluated when the data uploading of each IoT device can be supported by two widely-used IoT solutions, LB-IoT and UB-IoT. The relationship between the convergence performance and accuracy of FL model training and the impact of three key parameters, size of the dataset, batch size, and number of local training passes, has also been investigated. A joint optimization problem that can maximize the utilization of the computational resources at edge servers and also minimize the energy cost of the IoT network has been formulated. A distributed algorithm based on ADMM has been proposed to address the formulated problems. 
We have proved that the proposed algorithm 
will not cause any data leakage of local dataset and will preserve the private information related to the topological property of IoT network. Simulation results have been presented to show that the proposed framework can significantly reduce the cost with only a limited degradation in model convergence.

\vspace{-0.1in}
\section*{Acknowledgment}
This work was supported in part by the National Natural Science Foundation of China (NSFC) under Grant No. 62071193, the Key R \& D Program of Hubei Province of China under Grant No. 2020BAA002, and the U.S. National Science Foundation under Grant No. CCF-1908308.

\vspace{-0.1in}
\bibliography{DeepLearningRef}

\begin{thebibliography}{10}
\providecommand{\url}[1]{#1}
\csname url@samestyle\endcsname
\providecommand{\newblock}{\relax}
\providecommand{\bibinfo}[2]{#2}
\providecommand{\BIBentrySTDinterwordspacing}{\spaceskip=0pt\relax}
\providecommand{\BIBentryALTinterwordstretchfactor}{4}
\providecommand{\BIBentryALTinterwordspacing}{\spaceskip=\fontdimen2\font plus
\BIBentryALTinterwordstretchfactor\fontdimen3\font minus
  \fontdimen4\font\relax}
\providecommand{\BIBforeignlanguage}[2]{{%
\expandafter\ifx\csname l@#1\endcsname\relax
\typeout{** WARNING: IEEEtran.bst: No hyphenation pattern has been}%
\typeout{** loaded for the language `#1'. Using the pattern for}%
\typeout{** the default language instead.}%
\else
\language=\csname l@#1\endcsname
\fi
#2}}
\providecommand{\BIBdecl}{\relax}
\BIBdecl

\bibitem{Yu2018FogComputingIoT}
R.~{Yu}, G.~{Xue}, and X.~{Zhang}, ``Application provisioning in fog
  computing-enabled {I}nternet-of-{T}hings: A network perspective,'' in
  \emph{Proc. of IEEE INFOCOM}, Honolulu, HI, Apr. 2018.

\bibitem{Alameddine2019EdgeComputingIoTDyanmic}
H.~A. {Alameddine}, S.~{Sharafeddine}, S.~{Sebbah}, S.~{Ayoubi}, and C.~{Assi},
  ``Dynamic task offloading and scheduling for low-latency {I}o{T} services in
  multi-access edge computing,'' \emph{IEEE J. Sel. Area Commun.}, vol.~37,
  no.~3, pp. 668--682, Mar. 2019.

\bibitem{XY2018EHFogComputing}
Y.~Xiao and M.~Krunz, ``Dynamic network slicing for scalable fog computing
  systems with energy harvesting,'' \emph{IEEE J. Sel. Area Commun.}, vol.~36,
  no.~12, pp. 2640--2654, Dec. 2018.

\bibitem{Zhao2018FLNonIID}
Y.~Zhao, M.~Li, L.~Lai, N.~Suda, D.~Civin, and V.~Chandra, ``Federated learning
  with non-iid data,'' \emph{arXiv preprint 1806.00582}, Jun. 2018.

\bibitem{Wang2020FLNonIID}
H.~Wang, Z.~Kaplan, D.~Niu, and B.~Li, ``Optimizing federated learning on
  non-iid data with reinforcement learning,'' in \emph{Proc. of IEEE INFOCOM},
  Virtual Conference, Apr. 2020.

\bibitem{Tu2020FLFogCompNetwork}
Y.~Tu, Y.~Ruan, S.~Wagle, C.~G. Brinton, and C.~Joe-Wong, ``Network-aware
  optimization of distributed learning for fog computing,'' in \emph{Proc. of
  IEEE INFOCOM}, Virtual Conference, Apr. 2020.

\bibitem{McMahan2017FLfirstpaper}
B.~McMahan \emph{et~al.}, ``Communication-efficient learning of deep networks
  from decentralized data,'' in \emph{Proc. of the International Conference on
  Artificial Intelligence and Statistics}, Fort Lauderdale, FL, Apr. 2017, pp.
  1273--1282.

\bibitem{Bonawitz2019FLMassiveDeploy}
K.~Bonawitz \emph{et~al.}, ``Towards federated learning at scale: System
  design,'' \emph{arXiv preprint arXiv:1902.01046}, Mar. 2019.

\bibitem{deng2012mnist}
L.~Deng, ``The {MNIST} database of handwritten digit images for machine
  learning research,'' \emph{IEEE Signal Processing Magazine}, vol.~29, no.~6,
  pp. 141--142, Nov. 2012.

\bibitem{Bianchi1996CSMA}
G.~Bianchi, L.~Fratta, and M.~Oliveri, ``Performance evaluation and enhancement
  of the {CSMA/CA MAC} protocol for 802.11 wireless lans,'' in \emph{Proc. of
  PIMRC}, vol.~2, Taipei, Taiwan, Oct. 1996, pp. 392--396.

\bibitem{XY2018UnlicencedNetSlice}
Y.~Xiao and M.~Krunz, ``Distributed resource allocation for network slicing
  over licensed and unlicensed bands,'' \emph{IEEE J. Sel. Area Commun.},
  vol.~36, no.~10, pp. 2260 -- 2274, Dec. 2018.

\bibitem{Hirzallah2018DySPAN}
M.~{Hirzallah}, Y.~{Xiao}, and M.~{Krunz}, ``On modeling and optimizing
  {LTE/Wi-Fi} coexistence with prioritized traffic classes,'' in \emph{Proc. of
  IEEE DySPAN}, Seoul, South Korea, Oct. 2018, pp. 1--10.

\bibitem{ADMM}
S.~Boyd, N.~Parikh, E.~Chu, B.~Peleato, and J.~Eckstein, ``Distributed
  optimization and statistical learning via the alternating direction method of
  multipliers,'' \emph{Found. Trends Mach. Learn.}, vol.~3, no.~1, Jan. 2011.

\end{thebibliography}
\bibliographystyle{IEEEtran}

\end{document}